\documentclass[journal]{IEEEtran}

\usepackage{xcolor,soul,framed} %,caption

\colorlet{shadecolor}{yellow}
\usepackage[pdftex]{graphicx}

\graphicspath{{../pdf/}{../jpeg/}}
\DeclareGraphicsExtensions{.pdf,.jpeg,.png}
\usepackage{etoolbox,siunitx}
\usepackage[cmex10]{amsmath}
\usepackage{array}
\usepackage{mdwmath}
\usepackage{mdwtab}
\usepackage{eqparbox}
\usepackage{url}

\usepackage{cite}
\usepackage{booktabs} 
\usepackage{multirow}
\usepackage{multicol}
\usepackage{amsfonts}
\usepackage{float}
\usepackage{dsfont}

\usepackage{amssymb}% http://ctan.org/pkg/amssymb
\usepackage{pifont}% http://ctan.org/pkg/pifont
\newcommand{\cmark}{\ding{51}}%
\newcommand{\xmark}{\ding{55}}%

\usepackage{hyperref}
\usepackage{etoolbox}
\usepackage{siunitx}
\usepackage{scalerel}
\usepackage{comment}

\robustify\bfseries
\usepackage{balance}

\usepackage{algpseudocode}
\usepackage{algorithm}
\begin{document}
% \bstctlcite{IEEEexample:BSTcontrol}
    \title{NeuroHeed: Neuro-Steered Speaker Extraction \\ using EEG Signals}
    \author{Zexu~Pan,
    Marvin~Borsdorf, 
    Siqi~Cai, 
    Tanja~Schultz,~\IEEEmembership{Fellow,~IEEE},
    and~Haizhou~Li,~\IEEEmembership{Fellow,~IEEE}
    
    \thanks{This work is supported by 1) German Research Foundation (DFG) under Germany's Excellence Strategy (University Allowance, EXC 2077, University of Bremen); 2) National Natural Science Foundation of China (Grant No. 62271432); 3) Guangdong Provincial Key Laboratory of Big Data Computing, The Chinese University of Hong Kong, Shenzhen (Grant No. B10120210117-KP02); 4) Agency of Science, Technology and Research (A*STAR) under its AME Programmatic Funding Scheme (Project No. A18A2b0046). (\textit{Corresponding author: Siqi Cai}).}
    \thanks{Zexu Pan is with the Integrative Sciences and Engineering Programme, and the Institute of Data Science, National University of Singapore, 119077 Singapore (e-mail: pan\_zexu@u.nus.edu).}
    \thanks{Marvin Borsdorf is with the Machine Listening Lab, the University of Bremen, 28359 Bremen, Germany (e-mail: marvin.borsdorf@uni-bremen.de).}
    \thanks{Siqi Cai is with the Department of Electrical and Computer Engineering, National University of Singapore, 119077 Singapore (e-mail: elesiqi@nus.edu.sg).}
    \thanks{Tanja Schultz is with the Cognitive System Lab, the University of Bremen, 28359 Bremen, Germany (e-mail: tanja.schultz@uni-bremen.de).}
    \thanks{Haizhou Li is with the School of Data Science, The Chinese University of Hong Kong, Shenzhen, 518172 China, and the Machine Listening Lab, University of Bremen, 28359 Bremen, Germany (e-mail: haizhou.li@nus.edu.sg).}
    }
% The paper headers
% \markboth{IEEE/ACM TRANSACTIONS ON AUDIO, SPEECH, AND LANGUAGE PROCESSING
% }{Roberg \MakeLowercase{\textit{et al.}}: High-Efficiency Diode and Transistor Rectifiers}

% ====================================================================
\maketitle

\begin{abstract}
Humans possess the remarkable ability to selectively attend to a single speaker amidst competing voices and background noise, known as \textit{selective auditory attention}. Recent studies in auditory neuroscience indicate a strong correlation between the attended speech signal and the corresponding brain's elicited neuronal activities, which the latter can be measured using affordable and non-intrusive electroencephalography (EEG) devices. In this study, we present NeuroHeed, a speaker extraction model that leverages EEG signals to establish a neuronal attractor which is temporally associated with the speech stimulus, facilitating the extraction of the attended speech signal in a cocktail party scenario. We propose both an offline and an online NeuroHeed, with the latter designed for real-time inference. In the online NeuroHeed, we additionally propose an autoregressive speaker encoder, which accumulates past extracted speech signals for self-enrollment of the attended speaker information into an auditory attractor, that retains the attentional momentum over time. Online NeuroHeed extracts the current window of the speech signals with guidance from both attractors. Experimental results demonstrate that NeuroHeed effectively extracts brain-attended speech signals, achieving high signal quality, excellent perceptual quality, and intelligibility in a two-speaker scenario.
\end{abstract}

\begin{IEEEkeywords}
Selective auditory attention, EEG, speaker extraction, online, autoregressive, self-enrollment, attractor accumulation
\end{IEEEkeywords}

\IEEEpeerreviewmaketitle

\section{Introduction}
\label{sec:introduction}

\IEEEPARstart{T}{he} human brain possesses a remarkable capacity for speech comprehension, enabling individuals to selectively focus on a specific speaker's signal amidst a cacophony of other voices and background sounds, an ability commonly referred to as selective auditory attention in a ``cocktail party" scenario~\cite{bronkhorst2000cocktail,cherry1953some}. 
% This human aptitude relies on the intricate interplay between the auditory system and the cochlea. 
Consequently, the integration of smart hearing aids with selective auditory attention holds the promise of significantly reducing listening effort~\cite{wang2017deep}. Furthermore, this technology can serve as an essential frontend for various downstream speech applications, including active speaker detection~\cite{tao2021someone}, speech recognition~\cite{wang2022predict}, speaker verification~\cite{liu2022mfa}, speaker localization~\cite{qian2021multi}, and speech emotion recognition~\cite{pan2020multi}. The fundamental question, however, pertains to equipping machines with the same ability for selective auditory attention found in humans.

Speech separation (SS)~\cite{lyon1983computational,hershey2016deep,luo2019conv,luo2020dual,xu2018single,chen2017deep,zeghidour2020wavesplit,stoller2018wave,liu2019divide,wang2023tf} is a related research field that aims to separate a multi-talker speech signal into the individual source streams, for each speaker. With deep learning, exceptional progress has been made with models such as deep clustering~\cite{hershey2016deep}, Conv-TasNet~\cite{luo2019conv}, DPRNN~\cite{luo2020dual}, and TF-GridNet~\cite{wang2023tf}. SS networks perform best when the number of speakers in the mixture is known. However, one limitation is that the separated speech signals lack association with the listener's attention. Therefore, a subsequent post-processing network, such as utilizing brain electroencephalography (EEG) response~\cite{geravanchizadeh2021ear} or visual focus~\cite{pan2022seg}, is required to identify the attended speech track. Moreover, it is computationally inefficient to separate the speech of interfering speakers when it is not necessary for attention.

\begin{figure}
  \centering
  \includegraphics[width=0.95\linewidth]{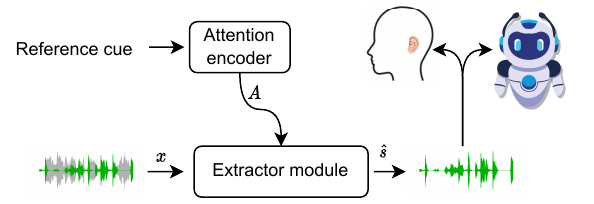}
  \caption{A typical speaker extraction model that extracts the attended speaker's voice according to an auxiliary reference cue. The extracted speech could serve as a live feed to the hearing-impaired, as well as for users and robots engaged in human-machine interactions.}
\label{fig:tse_network}
\end{figure}

Besides speech separation, speaker extraction (SE) is another approach to solve the ``cocktail party'' problem by extracting only the target speech signal from a multi-talker speech signal~\cite{Chenglin2020spex,spex_plus2020,wang2019voicefilter,he2020speakerfilter,vzmolikova2019speakerbeam,xiao2019single,shi2020speaker,delcroix2020improving,sato2021multimodal,ochiai2019multimodal,marvin2021,wang21aa_interspeech}. A typical SE model is illustrated in Fig.~\ref{fig:tse_network}, where an attention encoder encodes an auxiliary reference cue into an attractor $A$. This attractor is then utilized to guide the extraction of a target speech signal $\hat{s}$ from a given multi-talker speech signal $x$.

The use of reference cues in most SE models draws inspiration from the human attention mechanism. For example, when the brain pays attention to the voice of a familiar speaker, the reference cue could be a pre-enrolled speech signal~\cite{Chenglin2020spex, vzmolikova2019speakerbeam}. Alternatively, when a person focuses their attention on a speaker they are visually observing, the reference cue can be derived from the video sequence capturing the attended speaker's lip movement ~\cite{usev21,ephrat2018looking,wu2019time} or body motion~\cite{pan2022seg}. However, in complex real-world scenarios, the practicality of both acoustic and visual reference cues is limited. Pre-enrolling numerous speakers and selecting the right one to facilitate attention would be burdensome, and it may not be feasible for the listener to always visually track the speaker they are attending to.

We propose that the most accurate attention reference cue is the neuronal activity response in the brain when attending to speech signals. Auditory neuroscience studies have demonstrated a strong correlation between the attended speech signal and the elicited neuronal activities in the brain, enabling the decoding of auditory attention in a ``cocktail party'' scenario~\cite{mesgarani2012selective}, also known as auditory attention detection (AAD). Among the various cortical activity recording devices, EEG stands out as a non-invasive, cost-effective method with high temporal resolution. Several studies have validated the feasibility of detecting auditory attention in the human brain using EEG signals~\cite{choi2013quantifying,mirkovic2015decoding,cai2021low,cai2021eeg,wang2020robust,cai2021auditory}, which shines a light on the cognitive control of speech applications, such as hearing aids.

This paper introduces NeuroHeed, a neuro-steered speaker extraction model that utilizes the EEG signal as the sole auxiliary attention reference cue. NeuroHeed explores the temporal association between the EEG signals and the attended speech content by encoding the EEG signal into a frame-level neuronal attractor, which effectively guides a time-domain speaker extraction model. The experiments on KUL dataset~\cite{das2019auditory} demonstrate that NeuroHeed outperforms existing state-of-the-art models by a large margin in terms of signal quality, perceptual quality, and intelligibility in the two-speaker scenario.

Previous studies on SE and SS methods mostly operate in offline mode, where the multi-talker speech signal is processed at the utterance level. There is an increasing interest in online processing, which is required in practical applications~\cite{TEA_PSE2023,wang2020online,von2019all,han2019online,giri21_interspeech,wang2020voicefilter,han2023online,li2023design}. Most offline models can be directly evaluated in the online or causal mode, such as the DPRNN~\cite{luo2020dual}, where the processing of the current frame does not necessarily rely on future frames. However, if the temporal convolutional neural network (TCN)~\cite{luo2019conv} is part of the model, it is beneficial to change the TCN to causal TCN for the online mode.

In this paper, we also study the online implementation of NeuroHeed. Neuroscience studies~\cite{brungart2007cocktail,kidd2005advantage} have demonstrated that extended exposure to a speech stimulus leads to improved perceived listening quality, attributed to the gradual reinforcement of attention during the listening process. We are motivated to mimic this human ability in NeuroHeed. To this end, we propose an autoregressive mechanism that self-enrolls the past extracted speech into an auditory attractor, which preserves the momentum of auditory attention in NeuroHeed. To the best of our knowledge, this is the first work to self-enroll the attended speaker from past extracted speech during online inference. As a result, the online NeuroHeed relies on two attention cues, a) the temporal association between speech stimulus and the neuronal attractor encoded from the EEG signal, b) the attended speaker encoded by the accumulated auditory attractor. The experiments show that the self-enrolled momentum-preserved auditory attractor provides a remarkable performance gain for online inference.

% replicating the enhanced perception resulting from prolonged exposure to a speaker.

The remainder of the paper is structured as follows. In Section~\ref{sec:related_work}, we provide an overview of the related works in the field. In Section~\ref{sec:proposal}, we present the formulation of the proposed NeuroHeed model. In Section~\ref{sec:experimental_setup}, we describe the experimental setup. In Section~\ref{sec:results}, we report the results. Finally, Section~\ref{sec:conclusion} concludes the study.

\begin{figure*}
  \centering
  \includegraphics[width=0.95\linewidth]{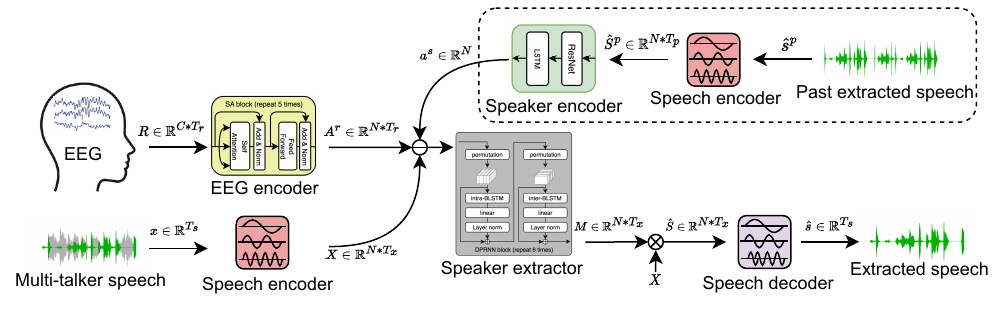}
  \caption{NeuroHeed: the proposed end-to-end neuro-steered speaker extraction model, that extracts the attended speaker's voice according to the attention detected from the auxiliary EEG signal. The modules in the dotted box are exclusive for online inference, which uses the past extracted speech as an additional reference cue. The symbol $\otimes$ represents element-wise multiplication, while the symbol $\ominus$ represents a concatenation operation along the channel dimension.}
  \label{fig:model}
\end{figure*}

\section{Related work}
\label{sec:related_work}
% In this section, we describe some related works.

\subsection{Speaker extraction with auditory attention attractor}
Humans can effortlessly recognize the speech signals of familiar speakers by their unique voice characteristics. Such voice characteristics can be easily obtained from a pre-enrolled speech signal and encoded into a fix-dimensional embedding vector, such as an i-vector~\cite{dehak2010front}, d-vector~\cite{wan2018generalized} or s-vector~\cite{svector2015}, through a speaker recognition model. The speaker embedding was used as the reference cue for speaker extraction (SE). % pre-enrolled speech reference cue for speaker extraction has been mostly studied.

The VoiceFilter~\cite{wang2019voicefilter} and TseNet~\cite{xu2019time} are examples of SE models that pre-train a speaker encoder to obtain the d-vector or i-vector as the auditory attractor. As such a speaker encoder is trained independently of the SE task, the derived auditory attractor may not be the best for speaker extraction. To address this problem, some propose to jointly train the speaker encoder and the speaker extraction model~\cite{delcroix2018single,vzmolikova2017learning,xu2019optimization}.

There have been studies to self-enroll the reference cues to avoid the speaker pre-enrollment~\cite{pan2020muse,pan2021reentry,afouras2019my,ito2021audio}. The reentry model~\cite{pan2020muse,pan2021reentry}
assumes that the visual reference is available and self-enrolls the auditory attractor from intermediate extracted speech, while self-enrollment for AVSE~\cite{ito2021audio} is performed from the mixture speech signals. However, none of the prior works utilizes past extracted speech for self-enrollment, and in real-time inference.

\subsection{Auditory attention detection from brain signals}

Neuroscience findings have revealed that at a ``cocktail party'', the attended speaker's voice correlates more to the listener's brain signals, than the competing background voices~\cite{parthasarathy2020bottomup}. Various brain signals, such as (ECoG)~\cite{mesgarani2012selective}, magnetoencephalography (MEG)~\cite{ding2012emergence}, and EEG~\cite{o2015attentional,biesmans2016auditory}, have been employed to decode auditory attention. Among them, EEG devices offer numerous advantages for brain-computer interface (BCI) applications, including wearability, non-invasiveness, and affordability. 

Studies have shown that it is possible to reconstruct the attended speech envelope from EEG signals to establish correlations with the attended speech stimulus~\cite{o2017neural,aroudi2020cognitive}. However, there is a trade-off between the accuracy and the decision window length. The AAD accuracy drops as the decision window narrows, which can be mainly attributed to the limited information of speech that is given in the low-frequency envelopes of small windows~\cite{geirnaert2021electroencephalography}. Recent advances have shown promising results by employing deep neural networks to directly decode brain attention using self-attention (SA)~\cite{su2022stanet,borsdorf2023multi} and cross-attention ~\cite{cai2021auditory}, which provides better accuracy in shorter decision windows. 

In most AAD studies, it is assumed that the uncorrupted single-speaker speech signals, usually referred to as clean speech, are given for both training and inference, and the recorded EEG signals are used to select one of the single-speaker speech as the attended one. However, in real-world conversational situations like ``cocktail party", clean single-speaker speech signals are typically not available, and correlating brain signals with the intended speaker's voice in the multi-talker speech signal is more challenging.

\subsection{Speaker extraction with neuronal attention attractor}
In view of the recent progress in EEG-based AAD and speech separation for multi-speaker speech mixture, some have proposed to separate the speech mixture into multiple single-speaker speech tracks, and select the speech track that has the highest correlation with the EEG signals as the attended speech signal~\cite{o2017neural,han2019speaker,o2015attentional,van2016eeg}. Unfortunately, this method usually requires the total number of speakers in the mixture signal to be known in advance, and, in addition, multi-talker speech separation unnecessarily leads to higher computational costs compared to the speaker extraction task. Each module in the cascaded pipeline is also subject to individual errors when not jointly optimized.

The brain-informed speech separation (BISS) model~\cite{biss2020} is motivated by the idea that the envelope of the attended speech signal can be reconstructed from the elicited EEG signal of a human listener. It uses this reconstructed envelope as the attention reference cue to perform speaker extraction in the frequency domain. The speech envelope reconstruction model is trained independently of the speaker extraction model.
The brain-enhanced speech denoiser (BESD) model~\cite{hosseini2021} and U-shaped BESD (UBESD) model~\cite{hosseini2022} take BISS a step further, which directly model the EEG signals with temporal convolutional neural network (TCN) and feature-wise linear modulation (FiLM)~\cite{perez2018film} to extract the neuronal attractor for a time-domain speaker extraction model, and show better performances. The state of the art is given by the brain-assisted speech enhancement network (BASEN)~\cite{zhang2023basen}, which is based on TCN for speech separation and EEG feature extraction. In addition, it applies a convolutional multi-layer cross-attention module for speech-EEG feature fusion. Both UBESD and BASEN up-sample the EEG signals to have the same time resolution as the speech features before feeding EEG signals into the EEG encoder, which unnecessarily increases the computational complexity.

Unlike UBESD and BASEN, our work is motivated by the self-attention (SA) models in recent AAD studies~\cite{borsdorf2023multi,cai2021auditory}. We propose to use SA layers in the EEG encoder to increase the receptive field (RF) for processing the EEG signals. To reinforce the ongoing attention and to improve the online inference, we make use of the past extracted speech via a self-enrollment mechanism. We do not up-sample the EEG signal to keep its computational complexity low. 
% Motivated by the recent dual-path recurrent neural network (DPRNN)~\cite{luo2020dual} for speech separation, we adopt DPRNN architecture for our proposed NeuroHeed to increase the RF for speech signals.

\section{NeuroHeed}
\label{sec:proposal}
In Fig.~\ref{fig:model}, we propose an end-to-end EEG-based neuro-steered speaker extraction model, called NeuroHeed. It consists of a speech encoder, a speaker extractor, a speech decoder, and two attention encoders, i.e., an EEG encoder and a speaker encoder.

%\subsection{Signal definition}

Let $x$ be a multi-talker speech signal in the time domain, which consists of the attended speech signal $s$ and the interfering speakers' speech signals $b_{i}$:
\begin{equation}
    \label{eqa:speaker_extraction}
    x = s + \sum_{i=1}^{I}b_{i}  \;  \in \mathbb{R}^{T_s}
\end{equation}
where $i \in \{1,...,I\}$ denotes the index of the interfering speaker. NeuroHeed seeks to extract $\hat{s}$ that is close to $s$ from $x$, conditioned on the EEG signal $R$ as the only auxiliary attention reference cue.

\subsection{Offline and online settings}
According to the mode of inference, we consider two settings, offline and online NeuroHeed. The offline NeuroHeed doesn't include the dotted box as shown in Fig.~\ref{fig:model} and it operates at the utterance level, in other words, the entire multi-talker speech signal $x$ is processed as input to calculate the output $\hat{s}$. The online NeuroHeed includes the dotted box as shown in Fig.~\ref{fig:model}, which processes a short time window ($w_c$) of input speech progressively, window by window. The processing of a speech window uses the past processed speech as the attention cue.
% , depending on the scenario (see Section~\ref{sec:on_off}). 
The latency of the online processing is usually the time window of $w_c$ if the required processing time is less than the time duration of $w_c$. 

In online mode, as illustrated in Fig.~\ref{fig:window}, a sliding window is applied on the multi-talker input speech signal $x$, which has two parts. The first part, denoted by $w_b$, is the past signal in a buffer that serves as the context for the processing of the current frame. The second part, denoted by $w_c$, is the current frame to be processed. Each time, the sliding window shifts by the length of $w_c$, and the data of the whole window length ($w_b + w_c$) is fed into NeuroHeed for processing.

\subsection{Model architecture}

\subsubsection{Speech encoder}
The speech encoder acts like a frequency analyzer to transform the $1$ dimensional (1D) multi-talker speech signal $x$ into a sequence of frame-based embeddings $X$ similar to the short-time Fourier transform (STFT). 

We adopt the time-domain approach~\cite{luo2019conv}, in which the speech encoder is a 1D convolutional layer (conv1D) followed by a rectified linear activation (ReLU):
\begin{equation}
    X = \text{ReLU}(\text{conv1D}(x,1, N, L))  \;  \in \mathbb{R}^{N\times T_x}
\end{equation}
where conv1D has input size $1$, output size $N$, kernel size $L$, and stride size $L/2$. The output $X$ is a $T_x$ frame embedding sequence of dimension $N$. $N$ and $L$ are set to $64$ and $20$, respectively, in this paper. 

The speech encoder is employed in both offline and online settings. In offline mode, the speech encoder is employed to encode the input multi-talker speech $x$. In online mode, it is also employed to encode the past-extracted speech $\hat{s}^p$ as illustrated in Fig. 2.

\begin{figure}
  \centering
  \includegraphics[width=0.6\linewidth]{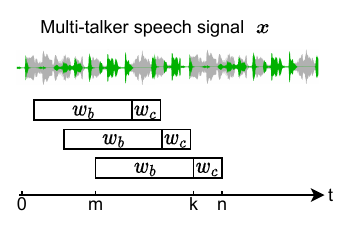}
  \vspace{-1.8mm} %To match the position of Figures 3 and 4
  \caption{In online mode, the multi-talker speech signal $x$ is processed via a sliding window that consists of two parts, i.e., $w_b$ for the buffered past signal and $w_c$ for the currently acquired signal to be processed. Each time, the sliding window shifts by the length of $w_c$.}
  \label{fig:window}
\end{figure}

\subsubsection{EEG encoder}
The EEG encoder aims to encode $C$-channel EEG signals $R$ into the neuronal attractor $A^r$, which provides a top-down guidance for speaker extraction. 

The EEG signals are known to be noisy, thus the longer the EEG signals, the better it correlates with the attended speech signals~\cite{o2017neural,aroudi2020cognitive}. In the previous studies, UBSED and BASEN models use a dilated TCN as the EEG encoder to increase the fixed-size receptive field. We are motivated by the recent AAD studies~\cite{borsdorf2023multi} to build the EEG encoder in NeuroHeed based on a global receptive field, which is designed as follows:
\begin{equation}
    A^r = \text{SA}(\text{PE}(\text{linear}(R, C, N)), N, N*4, 1, P_r)  \;  \in \mathbb{R}^{N\times T_r}
\end{equation}
which consists of a linear layer, a positional encoding (PE), and $P_r$ layers of SA~\cite{vaswani2017attention}. The linear layer has input size $C$ and output size $N$. Each SA has $1$ head,  hidden size $N$, and feed-forward size $N*4$. $C$ is set to $64$ abd $P_r$ is set to $5$ in this paper. The output $A^r$ is a $T_r$ frame embedding sequence of dimension $N$. The EEG encoder works in both offline and online settings. 

As a signal pre-processing, the raw EEG signals $R$ are first re-referenced to the average response of all electrodes. It is shown that non-linear AAD decoders benefit from broadband EEG information~\cite{de2020machine,vandecappelle2021eeg}. 
% Therefore, we further bandpass-filter the EEG signals between $1$ and $32$ Hz with a finite impulse response (FIR) and finally down-sample the signals to $128$ Hz.
Therefore, we further bandpass-filter the EEG signals between $1$ and $32$ Hz with a finite impulse response (FIR) and finally down-sample the signals from the recorded sampling rate of $8,192$ Hz to $128$ Hz. The frequency range is selected according to previous findings in the studies of a neural approach to AAD~\cite{su2022stanet,cai2021auditory}. 

The EEG encoder is trained together with the speaker extractor using the overall extracted speech quality as the learning objective, we expect that the EEG encoder learns to produce the EEG embedding sequence $A^r$ that associates with the attended speech temporally. In other words, we don't seek to explicitly reconstruct the envelope of the attended speech, but rather to generate EEG feature representations, that synchronize with the attended speech and reflect the attentive listening in the human brain.

\begin{figure}
  \centering
  \includegraphics[width=0.8\linewidth]{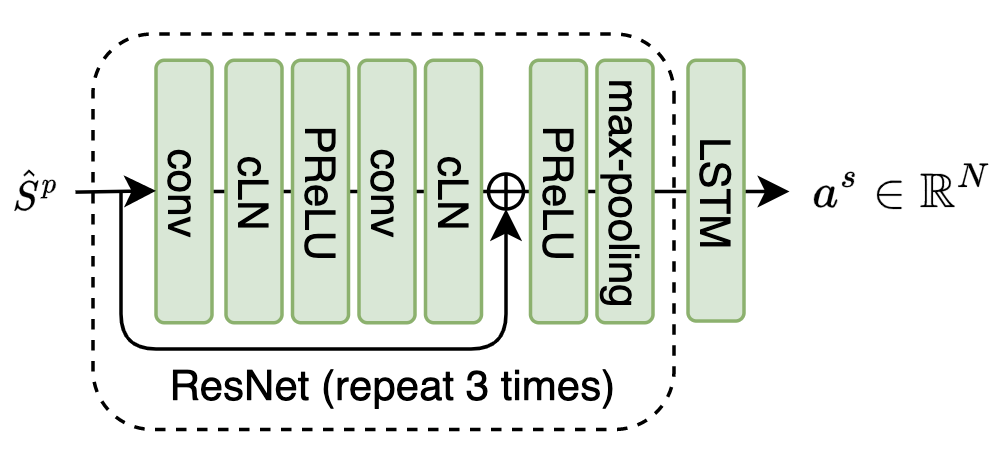}
  % \vspace*{-5mm}
  \caption{The model architecture of the speaker encoder, that consists of convolutional layers (conv), cumulative layer normalization (cLN)~\cite{luo2019conv}, parametric rectified linear activation (PReLU), max-pooling, and LSTM. The symbol $\oplus$ represents element-wise addition.}
  \label{fig:speaker_encoder}
\end{figure}

\subsubsection{Speaker encoder}
The speaker encoder is only applied in the online setting, where NeuroHeed performs sequential decoding with a sliding window as depicted in Fig.~\ref{fig:window}. We propose to encode the past extracted speech signal $\hat{s}^p$ into a fixed dimensional speaker embedding $a^s$. As $\hat{s}^p$ is single-speaker speech, it is representative of the recently attended speaker. We propose an autoregressive speaker encoder to perform such speaker self-enrollment. The resulting speaker embedding $a^s$ is also referred to as an auditory attractor, which is expected to maintain the continuity of auditory attention in the speaker extractor. 

%$a^s$ to explore the speaker characteristics of the attended speaker in processing the current window of speech signals. 

As in Fig.~\ref{fig:model}, the past extracted speech signal $\hat{s}^p$ is first processed by the speech encoder to derive its latent representation $\hat{S}^p$:
\begin{equation}
    \hat{S}^p = \text{ReLU}(\text{conv1D}(\hat{s}^p,1, N, L))  \;  \in \mathbb{R}^{N\times T_p}
\end{equation}
where it is a $T_p$ frame embedding sequence of dimension $N$.
Since $a^s$ serves as the top-down voluntary focus for speaker extraction, it is beneficial for $\hat{S}^p$ to share the same latent feature space with $X$. Thus, this speech encoder is the same speech encoder that processes the multi-talker speech signal~\cite{spex_plus2020}.

The speaker encoder consists of $3$ repeated ResNet~\cite{pan2021reentry,spex_plus2020} structures to extract the frame-level speaker representation from $\hat{S}^p$, which is further summarized into an embedding vector $a^s$ by a long short-term memory (LSTM) layer, as shown in Fig.~\ref{fig:speaker_encoder}

% \begin{equation}
%     a^s = \text{LSTM}(\text{ResNet}(\hat{S}^p))  \in \mathbb{R}^{N}
% \end{equation}

The online NeuroHeed is autoregressive: the past extracted speech is fed into the speaker encoder to update the auditory attractor $a^s$ for each sliding window, in order to extract $\hat{s}$. The attractor $a^s$ is taken from the last hidden state of the LSTM, thus it is an accumulation of all past attended speech signals. In offline mode, the speaker encoder module in NeuroHeed is excluded as the whole speech utterance is processed altogether.

\subsubsection{Speaker extractor}
The speaker extractor module aims to estimate a mask $M$ that only lets the attended speaker's voice pass in $X$, and the masked speech embedding $\hat{S}$ is obtained via:
\begin{equation}
    \hat{S} = X \otimes M  \;  \in \mathbb{R}^{N\times T_x}
\end{equation}
where $\otimes$ is element-wise multiplication.

Inspired by the success of SS models, where TCNs~\cite{luo2019conv} or dual-path recurrent neural networks (DPRNN)~\cite{luo2020dual} are used as speaker extractors to increase the RF for speech signal processing, we study both networks in NeuroHeed. Unlike SS, where a mask is estimated for every speaker, NeuroHeed only estimates the attended speaker's mask with guidance from a neuronal attractor $A^r$, or additionally with an auditory attractor $a^s$ for the online model. 

We took the concatenation approach to fuse the attention attractors with speech signals~\cite{Chenglin2020spex,pan2021reentry}. The neuronal attractor $A^r$ is a frame-based embedding that is temporally synchronized with the speech embedding $X$. We linearly interpolate $A^r$ from length $T_r$ to $T_x$ to match the frame rate before concatenation. Since the auditory attractor $a^s$ is an utterance-level embedding vector, it is repeated $T_x$ times for frame-level concatenation. For TCN-based speaker extractors, the attention attractors are concatenated with the speech embeddings at the start of every TCN block. For the DPRNN-based speaker extractor, they are concatenated only at the start of the speaker extractor. A linear projection layer is applied after the concatenation to map the embedding back to the original dimension before fusion.

\subsubsection{Speech decoder}
The speech decoder is an inverse operation of the speech encoder, that takes in the masked speech embedding $\hat{S}$ to reconstruct the time-domain waveform $\hat{s}$ through a linear layer and an Overlap-and-Add (OnA) operation~\cite{oppenheim1978theory}:
\begin{equation}
    \hat{s} = OnA(linear(\hat{S}, N, L), L/2) \in \mathbb{R}^{T_s}
\end{equation}
where the linear layer has the input size $N$ and output size $L$. The OnA operation has a frameshift of $L/2$.

\subsection{Objective function}

NeuroHeed is an end-to-end neural network that is trained to maximize the scale-invariant signal-to-noise ratio (SI-SDR)~\cite{le2019sdr} between the extracted speech signal $\hat{s}$ and the clean speech signal $s$. This is achieved by using the negative SI-SDR as an objective function during training as follows:
\begin{equation}
    \label{eqa:loss_sisnr}
    \mathcal{L}_{\text{SI-SDR}}(s, \hat{s}) = - 20 \log_{10} \frac{\big\|\frac{<\hat{s},s>}{\|s\|^2}s\big\|}{\big\|\hat{s} - \frac{<\hat{s},s>}{\|s\|^2}s\big\|}.
\end{equation}

The speech encoder, EEG encoder, speaker encoder, and speech decoder in NeuroHead are trained together with the speaker extractor using the overall extracted speech quality as the learning objective.

\subsection{Training of NeuroHeed for offline and online settings}
\label{sec:on_off}

In the offline setting, the autoregressive speaker encoder is not required by NeuroHeed. Thus, the training and inference processes for NeuroHeed are simple as the multi-talker speech signal $x$ is processed at the utterance level. 

However, in an online setting when the speaker encoder is involved, NeuroHeed is autoregressive, which may dramatically increase the computation during training if the actual past extracted speech is used for every sliding window. In the recurrent neural network (RNN) literature for tasks like machine translation (MT) or natural language processing (NLP), teacher forcing~\cite{williams1989learning} is proposed to use ground truth data as input instead of past model outputs in training. To overcome the mismatch between ground truth data during training and predicted data during inference, also known as exposure bias, scheduled sampling~\cite{bengio2015scheduled} is proposed in training to gradually minimize the mismatch, forcing the model to deal with its own mistakes.

A key difference between SE and the aforementioned NLP tasks is that SE can also work without any past model output independent of offline or online inference. Therefore, to cope with the challenges in autoregressive training and based on the characteristics of speaker extraction, we design a two-pass speaker encoder dropout training for the online NeuroHeed.

\textbf{Speaker encoder dropout:} We first design the online NeuroHeed to work with or without the speaker encoder module with a dropout mechanism. This is necessary because there are cases in practice where no past extracted speech is accessible. With a probability $\mathds{P}(\text{dropout})=0.2$, we drop out the speaker encoder module and set $a^s$ to be a zero vector, such that the model learns not to rely on $a^s$ when it is unavailable.

\textbf{Two-pass:} When the above-mentioned speaker encoder dropout is not applied, the model relies on past extracted speech $\hat{s}^p$ to obtain an $a^s$. As it is time-consuming to obtain the actual $\hat{s}^p$ during training, we introduce a pseudo past extracted speech to replace the use of $\hat{s}^p$. 

Taking an example from Fig.~\ref{fig:window}, during training, for each forward propagation of an utterance $x$, we train the model on a random segment $x[m:n]$. NeuroHeed is run twice for each forward propagation. The first run takes in past multi-talker speech signals $x[0:k]$ and the corresponding EEG signal $R[0:k']$ as input, and outputs the pseudo past extracted speech $\hat{s}[0:k]$, without the use of the speaker encoder. The second pass takes in $x[m:n]$, $R[m':n']$, and $\hat{s}[0:k]$ as input, and outputs $\hat{s}[m:n]$, using the complete NeuroHeed. The forward propagation in training is shown in Algorithm~\ref{alg:train}.

\begin{algorithm} [t]
\caption{The pseudo-code for a forward propagation to train the autoregressive online NeuroHeed.}\label{alg:train}
\begin{algorithmic}

\State Input: $x \in \mathbb{R}^{T_s}$
\State Input: $R \in \mathbb{R}^{N\times T_r}$     %\Comment{This is a comment}
\\
% \State Optimizer.zeroGradient()
% \\
\State \# Randomly set window length 
\State n - k = random.choice([0.05, 0.1, 0.2]) 
\State k - m = random.uniform(1.0,10.0)
\State m = random.randint(0, $T_s$ - (n-m))
\\
\If{random.uniform(0,1) $<$ 0.2}:
    \State \# Speaker encoder dropout
    \State $\hat{s}[m:n]$ = NeuroHeed($x[m:n]$, $R[m':n']$)
\Else:
    \State \# two-pass training with pseudo past extracted speech
    \State $\hat{s}[0:k]$ = NeuroHeed($x[0:k]$, $R[0:k']$)
    \State $\hat{s}[m:n]$ = NeuroHeed($x[m:n]$, $R[m':n']$, $\hat{s}[0:k]$)
\EndIf
\\
\State loss = $\mathcal{L}_{\text{SI-SDR}}(s[m:n], \hat{s}[m:n])$
\\
\State loss.backward()
\State Optimizer.step()
\State Optimizer.zeroGradient()

\end{algorithmic}
\end{algorithm}

Although the pseudo past extracted speech is used during training, the actual past extracted speech $\hat{s}^p$ is used for speaker encoder during inference. We expect little mismatch as both the pseudo and actual $\hat{s}^p$ are real NeuroHeed outputs. It is also worth noting that the proposed two-pass dropout training mechanism for autoregressive online speaker extraction models is independent of the modality of the auxiliary cue, meaning that it could also work for other auxiliary references such as lip recording instead of EEG signals.

\subsection{Inference normalization for online NeuroHeed}
\label{sec:norm}
The negative SI-SDR objective function is scale-invariant, so the energy level of the NeuroHeed output is not bounded, and the energy level of the sliding windows may vary from window to window. Normalization of the current output over the output history is needed to keep a consistent energy level for intelligibility. We propose an inference normalization (Inf. Norm), that compares the energy between the buffered output signals $\hat{s}[m:k]$ and the past extracted speech $\hat{s}^p[m:k]$:
\begin{equation}
    \hat{s}[m:n] = \hat{s}[m:n] * \frac{||\hat{s}^p[m:n]||}{||\hat{s}[m:n]||}
\end{equation}

Since the first few sliding windows have a relatively short duration, the energy levels are not representative. To stabilize the energy normalization, we process the first second of the mixture speech as a whole instead of using sliding windows, which means that the model initialization takes $1$ second in practical implementation (1s init).

Other scale-sensitive loss functions such as signal-to-distortion ratio (SDR) could also be used. However, we experimentally found out that using SI-SDR leads to better overall signal quality.

\section{Experimental setup}
\label{sec:experimental_setup}

\subsection{Dataset}
\label{sec:dataset}
We evaluate our proposed model on the publicly available KULeuven (KUL) dataset~\cite{das2019auditory} which contains EEG signals collected from $16$ normal-hearing subjects using the BioSemi ActiveTwo system. The EEG signal is recorded at a $8,192$ Hz sampling rate with $64$ channels. In total, $20$ trials are collected for each subject. In our experiments, we used the first eight trials, such that the subjects are not attending to the same speech stimulus repeatedly. In each trial, subjects are instructed to listen to the speech perceived by one ear and ignore the speech that enters the other ear, when using plugged-in earphones. The speech signals are based on four Dutch short stories narrated by two male speakers which were played to all subjects. To ensure that the subjects were focusing on the target speech, they were asked to complete a set of multiple-choice questions after each trial. 
% More details of the KULeuven dataset can be found in~\cite{vandecappelle2021eeg,das2019auditory}

Since there are only two speakers given in the KUL dataset as speech stimuli, we perform subject-dependent and speaker-dependent studies. For each trial, we randomly split the data into the training, validation, and test sets with a ratio of $75\%$, $12.5\%$, and $12.5\%$, respectively, without overlap of speech stimulus between sets. For training and validation, we randomly segment the speech signals into lengths varying from $1$ to $10$ seconds to fit into the memory of the graphics processing unit (GPU), while the lengths in the test set vary from $1$ to $15$ seconds. We allow mixtures signals to have overlapping regions with each other to increase the total number of samples. 

We also perform a mixture signal augmentation (Train Aug.), in which the target speech signal is mixed with a speech signal randomly selected from another part of the story, instead of mixed with the actual interfering speech signal. We apply a random signal-to-noise (SNR) ratio between $10$ dB to $-10$ dB, only during training. In total, we have $400,000$ utterances for training, $3,000$ utterances for validation, and $3,000$ utterances for testing. The speech signals are sampled at 8 kHz in this study.

\subsection{Models}

\subsubsection{PIT-separation upper bound}
We build a speech separation model that is based on permutation invariant training (PIT)~\cite{kolbaek2017multitalker}, called PIT-separation, using DPRNN~\cite{luo2020dual}. Different from NeuroHeed, PIT-separation performs speech separation. Instead of relying on EEG signals to find the target speech, PIT-separation assigns one of the separated speech signals to the target speaker based on the permutation that provides the highest SI-SDR value. Therefore, this model can be considered as an upper bound for NeuroHeed.

\subsubsection{BISS baseline}
The brain-informed speech separation (BISS)~\cite{biss2020} is the pioneering work that directly extracts the attended speech signal based on the EEG signal. First, it reconstructs the attended speech envelope from the EEG signal by applying an EEG-to-envelope regression model. Next, it uses the envelope in a frequency-domain two-dimensional TCN (2D-TCN) network for speaker extraction. We re-implement the BISS model and report its results here. Besides BISS, we also build a model named BISS$^{+}$, in which we replace the EEG-to-envelope regression model with our proposed SA as the EEG encoder.

\subsubsection{UBESD baseline}
The U-shaped brain-enhanced speech denoiser (UBESD) model~\cite{hosseini2021,hosseini2022} is an end-to-end neural network that uses the EEG signal to denoise a multi-talker speech signal in the time domain. It applies TCN to construct an EEG encoder as well as a U-shaped speaker separator (U-TCN), and uses skip connections at the layers which have the same resolution as in the U-shaped architecture. It also uses a Feature-wise Linear Modulation (FiLM) between the EEG features and the speech features to extract the EEG features of the attended speaker and for fusion at multiple stages. The EEG signal is first processed to have the same resolution as the speech feature before feeding it into the EEG encoder.  Unlike NeuroHeed, UBESD directly estimates the target signal instead of estimating a mask for the target speaker. We re-implement the UBESD model and report its results here. Besides UBESD, we also construct a model named UBESD$^{+}$, that replaces the TCN and FiLM modules with our proposed SA as the EEG encoder.

\subsubsection{BASEN baseline}
The brain-assisted speech enhancement network (BASEN)~\cite{zhang2023basen} is the current state-of-the-art EEG-based neuro-steered speaker extraction model. It adopts a TCN architecture for both the EEG encoder and the speaker extractor module. It also proposes a convolutional multi-layer cross-attention (CMCA) module to fuse the EEG features with the speech features. The EEG signal is first processed to have the same resolution as the speech feature before feeding into the EEG encoder. We re-implemented the BASEN model and report its results here. Besides BASEN, we also build a model named BASEN$^{+}$, that replaces the TCN modules with our proposed SA as the EEG encoder.

\subsubsection{NeuroHeed ablations}
% The offline NeuroHeed differs from the UBESD baseline in two aspects, namely the DPRNN-based speaker extractor and the SA-based EEG encoder. We alter the network architectures of NeuroHeed to study the contribution from differences. We replace the DPRNN with U-DNCC in NeuroHeed to form NeuroHeed$^{+}$. We replace the SA with FiLM+TCN in NeuroHeed to form NeuroHeed$^{\mathparagraph}$. 

Besides the DPRNN architecture, we additionally study a speaker extractor based on TCN as this is also widely used in the speech separation and speaker extraction literature. We replace the DPRNN with TCN to form NeuroHeed$^{\dagger}$. We also study causal-TCN, in which the model is trained to not rely on future frames when the current frame is processed, referred to as NeuroHeed$^{\ddagger}$. This model could be used in online inference.

\begin{table*}
    \centering
    \sisetup{
    detect-weight, % Make siunitx detect align bold cells correctly
    mode=text, % Make siuntix print tables in text mode (causes width of bold characters to be the same as non-bold)
    tight-spacing=true,
    round-mode=places,
    round-precision=2
    }
    \caption{Offline evaluation for EEG-based neuro-steered speaker extraction models, with different EEG encoders, loss functions, and with (\cmark) or without (\xmark) training augmentation (Train Aug.). SI-SDRi and SDRi are reported in dB. The percentage positive rate (PPR) is the percentage of the extracted speech in the test set that has both a positive SI-SDRi value and a higher SI-SDRi value with respect to the attended speech than to the interfering speech. The number of parameters (Param) are reported in million. Note that in offline models, the speaker encoder is not used at all.}
    % \addtolength{\tabcolsep}{-1.5pt}
    \resizebox{\linewidth}{!}{
    \begin{tabular}{c|c|c|c|c|c|c|c|c|c|c|c} 
       \toprule
        Sys. \#     &Model   &Speaker extractor &EEG encoder &Loss &Train Aug. &SI-SDRi   &SDRi   &PESQi  &STOIi & PPR   &Param\\
        \midrule  
        0   &PIT-separation   &DPRNN &-    &SI-SDR &\cmark
            &19.4  &19.6  &1.90  &0.23  &100.0\%      &2.6 \\
            \midrule  
        1   &BISS~\cite{biss2020}   &2D-TCN &Envelope &\multirow{3}{*}{SI-SDR} &\multirow{3}{*}{\cmark}
            &-0.1	&0.5	&-0.08	&-0.03	&59.4\%    &0.5  \\
        2   &UBESD~\cite{hosseini2022}  &U-TCN  &TCN+FiLM &&
            &5.1   &5.8   &0.09  &0.03  &80.9\% &2.4    \\
        3   &BASEN~\cite{zhang2023basen}    &TCN    &TCN+CMCA &&
            &5.6   &6.7   &0.22  &0.03  &75.6\% &0.7    \\
        \midrule  
        4   &BISS$^{+}$   &2D-TCN  &SA &\multirow{3}{*}{SI-SDR}&\multirow{3}{*}{\cmark}
            &8.4   &8.9   &0.44  &0.10  &89.6\%    &0.8\\
        5   &UBESD$^{+}$   &U-TCN   &SA & &
            &6.1   &7.0   &0.17  &0.04  &85.2\% &2.0\\
        6   &BASEN$^{+}$   &TCN  &SA+CMCA &&
            &11.5  &12.4  &0.59  &0.13  &91.0\% &0.7\\
        \midrule
        7   &NeuroHeed$^{\dagger}$ &TCN   &\multirow{3}{*}{SA} &\multirow{3}{*}{SI-SDR} &\multirow{3}{*}{\cmark}
            &\textbf{15.0}  &\textbf{16.0}  &\textbf{1.08}  &\textbf{0.18}  &91.5\% &10.6     \\
        8   &NeuroHeed$^{\ddagger}$   &causal-TCN  & &&
            &10.5	&11.1	&0.58	&0.13	&89.4\% &10.6     \\
        9   &NeuroHeed  &DPRNN   & &&
            &14.3  &15.5  &0.95  &0.16  &90.8\% &2.9     \\
        \midrule
        10   &\multirow{2}{*}{NeuroHeed}  &\multirow{2}{*}{DPRNN}   &\multirow{2}{*}{SA} &SI-SDR    &\xmark 
            &8.3   &9.2   &0.26  &0.09  &\textbf{91.6\%} &\multirow{2}{*}{2.9}\\
        11  &&& &SDR    &\cmark 
                &12.6  &14.7  &0.89  &0.13  &88.6\%   &\\
        \bottomrule
    \end{tabular}
    }
    % \addtolength{\tabcolsep}{1.5pt}
    % \vspace*{-3mm}
    \label{tab:off_line}
\end{table*}

\subsection{Training details}
For the model implementation and to conduct our experiments, we use PyTorch\footnote{The data preparation and NeuroHeed training codes are available at https://github.com/zexupan/NeuroHeed.} as framework. During training. we apply Adam~\cite{kingma2015adam} optimizer and increase the learning rate (lr) for the first $warmup_n = 15,000$ training steps as follows:
\begin{equation}
    \text{lr} = 0.1 \cdot 64^{-0.5} \cdot step_n \cdot warmup_n^{-1.5}
\end{equation}
where $step_n$ is the step number. After the warmup is done, the learning rate is halved when the best validation loss does not improve within 6 consecutive epochs. The training stops when the best validation loss does not improve within 10 subsequent epochs. NeuroHeed is trained on four Tesla \mbox{32 GB} V100 GPUs, with an effective batch size of $32$.

We found that the NeuroHeed$^{\dagger}$ which adopts TCN as the speaker extractor is hard to converge when training from scratch. Thus, its EEG encoder is initialized with the trained EEG encoder from NeuroHeed.

When training the online NeuroHeed, the speech encoder, speech decoder, EEG encoder, and speaker extractor are initialized with the weights from the offline NeuroHeed to speed up the model's convergence and to reduce the training time. 

\subsection{Evaluation metrics}
To evaluate the extracted speech, we use the improvement in SI-SDR (SI-SDRi)~\cite{le2019sdr}; the improvement in SDR (SDRi)~\cite{vincent2006performance}; the improvement in perceptual evaluation of speech quality (PESQi)~\cite{rix2001perceptual}, and the improvement in Short Term Objective Intelligibility  (STOIi)~\cite{taal2010short}. All improvements are calculated with respect to the unprocessed multi-talker speech signals. Both the SI-SDRi and the SDRi represent signal quality while PESQi and STOIi represent perceptual quality and intelligibility, respectively. The higher the better for all metrics.

In addition, we report the percentage positive rate (PPR), which is defined as the percentage of extracted speech signals that satisfies both i) a positive SI-SDRi value, ii) a higher SI-SDRi value with respect to the target speech signal than to the interfering speech signal. As sometimes the SI-SDR of the extracted speech with both the target and interfering speakers may be low, thus satisfying both criteria does not necessarily represent the model's ability to extract the correct speakers, but it gives a rough estimate and the higher the better.

In online evaluation, we use a real-time factor (RTF) to express the model's computational complexity, the higher the better. It is calculated using one 1080Ti GPU.

\begin{figure*}
  \centering
  \includegraphics[width=0.99\linewidth]{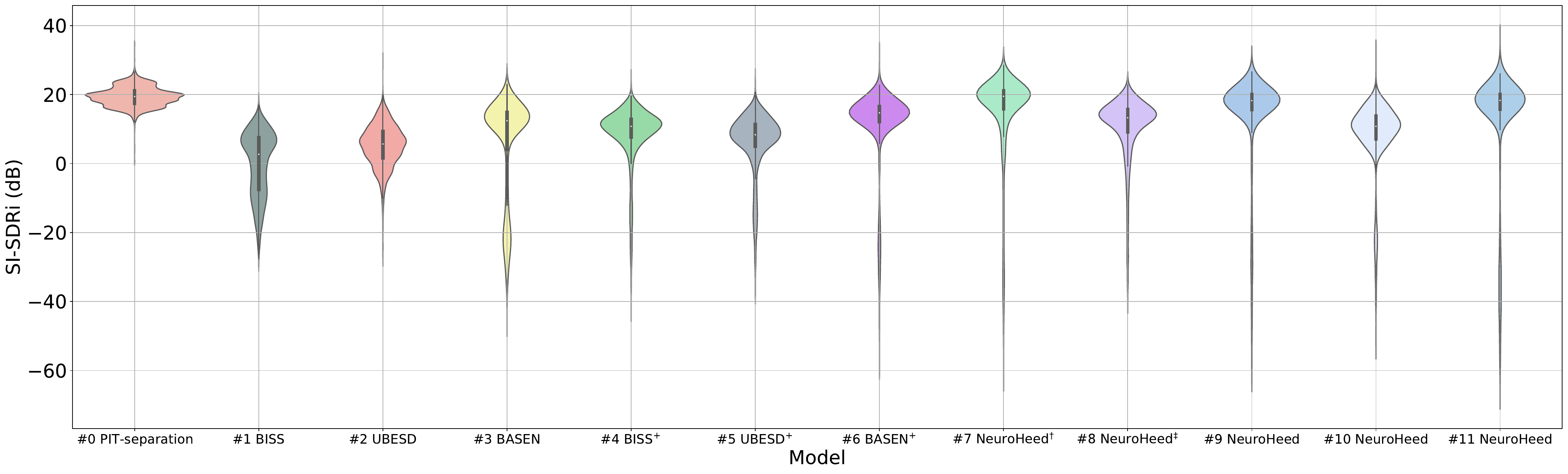}
  \caption{Violin plot of extracted speech signal's SI-SDRi from the test set for various models in offline evaluation.}
  \label{fig:offline_violin}
\end{figure*}

\begin{figure}
  \centering
  \includegraphics[width=0.99\linewidth]{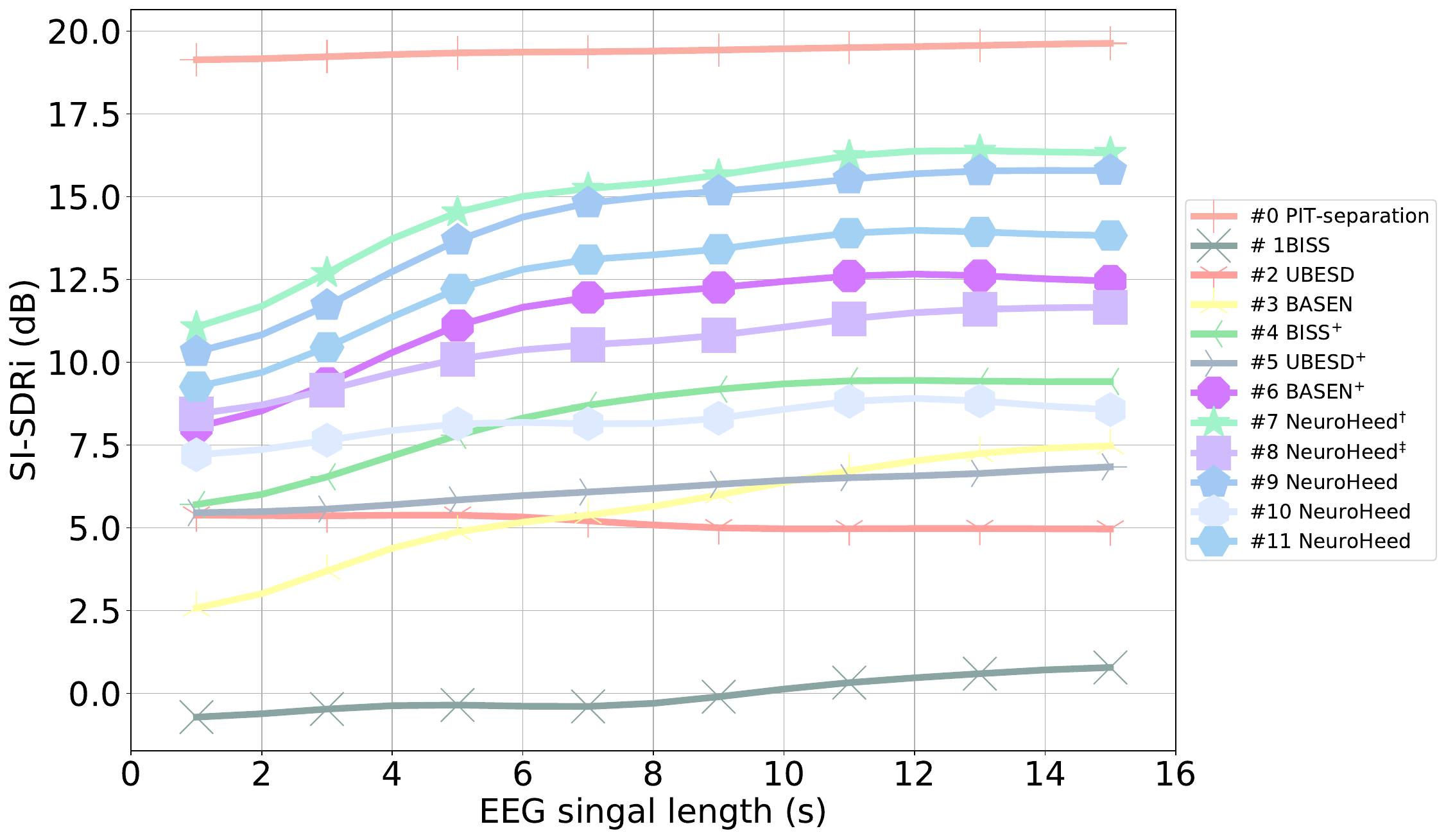}
  \caption{The average SI-SDRi for various utterance lengths in the offline evaluation. Longer utterances mean longer EEG signals associated.}
  \label{fig:offline_time}
\end{figure}

\begin{figure*}
  \centering
  \includegraphics[width=0.99\linewidth]{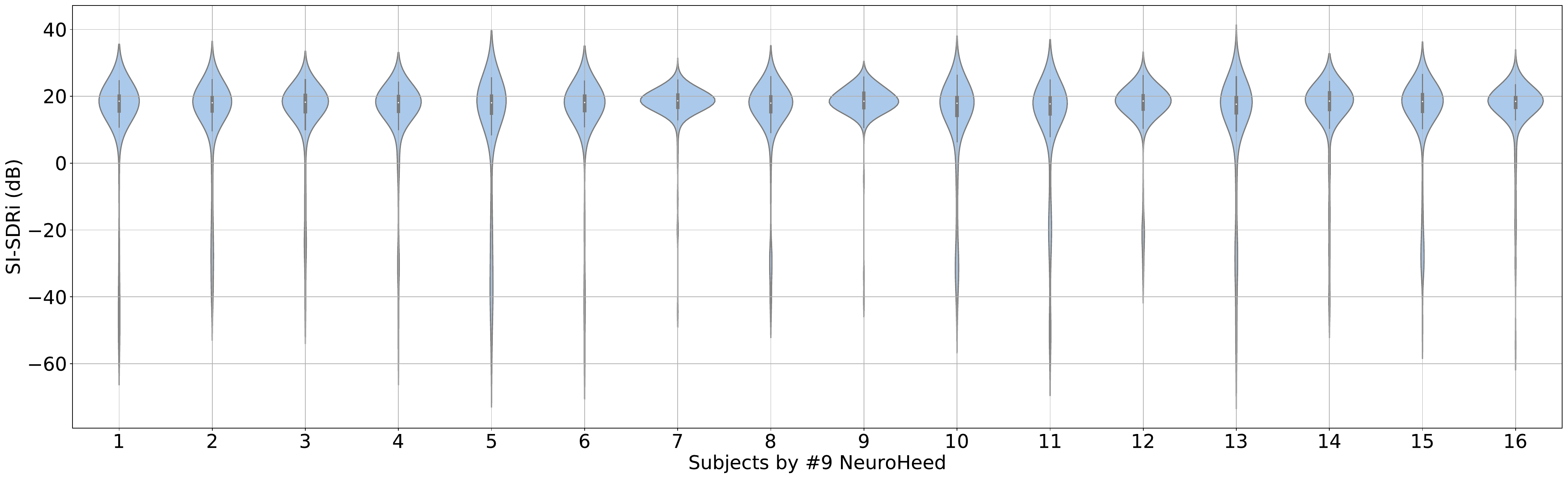}
  \caption{Violin plot of extracted speech signal’s SI-SDRi in the offline evaluation by subjects for system \#9 NeuroHeed.}
  \label{fig:offline_subject}
\end{figure*}

\begin{figure*}
  \centering
  \includegraphics[width=0.99\linewidth]{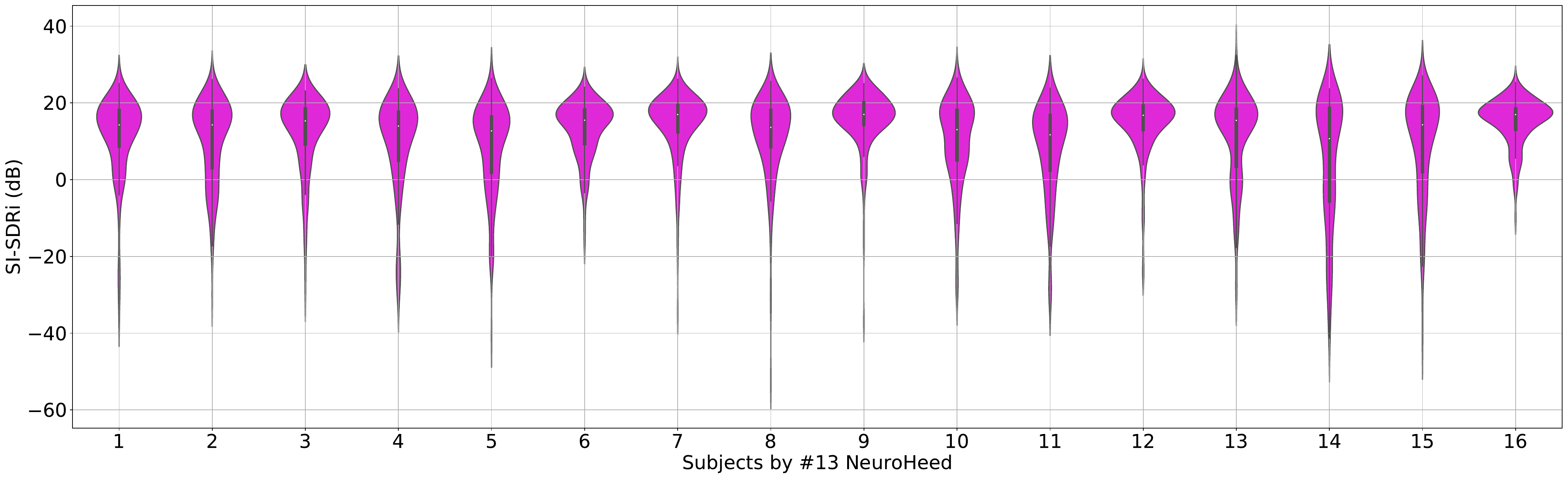}
  \caption{Violin plot of extracted speech signal’s SI-SDRi in the online evaluation by subjects for system \#13 NeuroHeed.}
  \label{fig:online_subject}
\end{figure*}

\section{Results}
\label{sec:results}
We report the results of the offline evaluation in section~\ref{sec:offline} and for the online evaluation in section~\ref{sec:online}.
We assign a system number (\#) to each trained model. Some models in the Table may appear to have the same system number because they are the same model but are evaluated differently, e.g., with or without the inference normalization, or online versus offline inference. 

\subsection{Offline evaluation}
\label{sec:offline}

\subsubsection{Comparison with baselines}
In Table~\ref{tab:off_line}, PIT-separation (\#0) shows the best results across all metrics with an incredibly high average SI-SDRi of $19.4$ dB. This is because it performs speech separation and optimal target speaker selection without being affected by an EEG reference cue. It has a percentage positive rate of $100\%$. In Fig.~\ref{fig:offline_violin}, we show the violin plot of SI-SDRi from all utterances in the test set. It can be seen that almost all utterances from the PIT-separation model have a SI-SDRi distributed around $20$ dB. The PIT-separation model sets the upper bound for other models in offline evaluation. In addition, shown in Fig.~\ref{fig:offline_time}, it is observed that PIT-separation performs consistently well for various utterance lengths.

In Table~\ref{tab:off_line}, our proposed NeuroHeed (\#9) outperforms all baseline models BISS, UBESD, and BASEN (\#1-3) by a large margin on the average SI-SDRi, SDRi, PESQi, STOIi, and PPR, showing that NeuroHeed is able to extract the target speech with better signal quality, perceptual quality, and intelligibility. The violin plot in Fig.~\ref{fig:offline_violin} shows that our proposed NeuroHeed has the SI-SDRi distribution centered around $18$ dB with low variance, while the baselines have the SI-SDRi distribution centered at a lower value with larger variance. In addition, we show the violin plot of the SI-SDRi from individual subjects for NeuroHeed (\#9) in Fig.~\ref{fig:offline_subject}. It can be seen that for almost all subjects the SI-SDRi distribution is centered at around $18$ dB, indicating a stable performance. In Fig.~\ref{fig:offline_time}, the NeuroHeed also has higher SI-SDRi for various utterance lengths compared to the baselines. 

Our proposed NeuroHeed differs from the baselines with respect to the speaker extractor and the EEG encoder. In systems \#4-5, we replace the EEG encoder of the baseline models by SA. In Table~\ref{tab:off_line}, it can be seen that BISS$^{+}$ has an improvement in SI-SDRi of $8.5$ dB compared to BISS, the SI-SDRi of UBESD$^{+}$ is $1.0$ dB higher compared to UBESD, and the BASEN$^{+}$ shows an improvement in SI-SDRi of $5.9$ dB compared to BASEN. The violin plot in Fig.~\ref{fig:offline_violin} shows that the BISS$^{+}$ has the SI-SDRi distribution centered at a higher value with smaller variance compared to BISS, the UBESD$^{+}$ has the SI-SDRi distribution centered at a higher value with smaller variance compared to UBESD, and the SI-SDRi distribution for the BASEN$^{+}$ has smaller variance compared to BASEN and less samples with very low SI-SDRi, showing the effectiveness of applying SA for modeling the EEG signals. In Fig.~\ref{fig:offline_time}, it can be seen that BISS$^{+}$, UBESD$^{+}$ and BASEN$^{+}$ outperform BISS, UBESD and BASEN, respectively, for various utterance lengths.

\subsubsection{NeuroHeed with different speaker extractor}

In Table~\ref{tab:off_line}, NeuroHeed with DPRNN as speaker extractor (\#9) performs comparably with NeuroHeed$^{\dagger}$ (\#7) in which TCN is used as speaker extractor, but comprises much fewer model parameters. NeuroHeed can be directly deployed online but hard for NeuroHeed$^{\dagger}$, as the latter needs to take into account future frames when processing the current frame. NeuroHeed$^{\ddagger}$ uses a causal TCN speaker extractor and can be deployed online. It has the same receptive field as NeuroHeed$^{\dagger}$, but its performance degrades sharply, with the average SI-SDRi dropping from $15.0$ dB to $10.5$ dB. 
In Fig.~\ref{fig:offline_violin}, it can be seen that the systems \#7-9 have some extracted speech signals with very negative SI-SDRi at around $-40$ dB. This may be caused by the network extracting the wrong speakers, resulting in very negative SI-SDRi values.

In Fig.~\ref{fig:offline_time}, we plot the average SI-SDRi as a function of the test utterance length. NeuroHeed$^{\dagger}$ and NeuroHeed$^{\ddagger}$ have a fixed speech receptive field in the TCN-based speaker extractor, but the average SI-SDRi still benefits from longer utterances. This is because the SA-based EEG encoder actually implicitly embraces a global receptive field covering a longer range. The same upward trend is observed for NeuroHeed.

\begin{table*}[t]
    \centering
    % \sisetup{
    % detect-weight, % Make siunitx detect align bold cells correctly
    % mode=text, % Make siuntix print tables in text mode (causes width of bold characters to be the same as non-bold)
    % tight-spacing=true,
    % round-mode=places,
    % round-precision=2,
    % table-format=2.2
    % }
    \caption{Online evaluation for EEG-based neuro-steered speaker extraction models. SI-SDRi and SDRi are reported in dB. The models reported in this Table use the online sliding window with $w_b$ of 2.5 seconds length and $w_c$ of 0.1 seconds length. Inference normalization (Inf. Norm) and $1$ second initialization ($1$s init) introduced for online evaluation in Sec~\ref{sec:norm} are also studied here.}
    \addtolength{\tabcolsep}{-2pt}
    % \resizebox{\linewidth}{!}{
    \begin{tabular}{c|c|c|c|c|c|c|c|c|c|c|c|c}
       \toprule
        Sys. \#     &Model &Speaker extractor   &Speaker encoder &Loss &Inf. Norm  &$1$s init &SI-SDRi   &SDRi   &PESQi  &STOIi & PPR  &RTF\\
        \midrule
        8   &NeuroHeed$^{\ddagger}$    &causal-TCN &\multirow{2}{*}{\xmark} &\multirow{2}{*}{SI-SDR}    &\multirow{2}{*}{\cmark}    &\multirow{2}{*}{\cmark}
            &7.8	&8.4	&0.42	&0.11	&84.2\%	  &2.0\\
        9   &NeuroHeed  &DPRNN &  &&&
            &8.3	&8.8	&0.49	&0.10	&78.0\%   &\textbf{4.5}\\
            
        \midrule    
        12   &NeuroHeed$^{\ddagger}$    &causal-TCN &\multirow{2}{*}{\cmark}  &\multirow{2}{*}{SI-SDR}   &\multirow{2}{*}{\cmark}    &\multirow{2}{*}{\cmark}
            &10.1  &11.0	&0.64	&0.12	&\textbf{86.2\%}	   &1.8\\
        13  &NeuroHeed  &DPRNN &  &&&
            &\textbf{11.2}  &\textbf{11.8}  &\textbf{0.69}  &\textbf{0.13}  &85.1\%    &3.6\\
        \midrule
        13*   &\multirow{3}{*}{NeuroHeed}  &\multirow{3}{*}{DPRNN} &\multirow{3}{*}{\cmark}  &SI-SDR &\cmark &\xmark   
            &9.7   &10.1  &0.61  &0.12  &84.5\%     &\multirow{3}{*}{3.6}\\
        13**   &&& &SI-SDR &\xmark &\cmark 
            &11.0  &11.6  &0.68  &0.13  &84.3\%   &\\
        14   &&& &SDR &\xmark &\cmark 
            &8.4   &9.2   &0.54  &0.10  &81.3\%     &\\
        \bottomrule
    \end{tabular}
    % }
    \addtolength{\tabcolsep}{2pt}
    % \vspace*{-3mm}
    \label{tab:on_line}
\end{table*}

\subsubsection{Ablation studies}

As described in section~\ref{sec:dataset}, we propose a mixture signal augmentation for the training dataset that replaces the actual interfering speech with another irrelevant interfering speech from another random part of a story. Since the actual interfering signal is also weakly correlated to the EEG signal, it is unclear whether the augmentation helps. Therefore, we present system \#10 which is the same as system \#9, and train it on the same amount of data, except that we do not perform the augmentation. As depicted in Table~\ref{tab:off_line}, the performance drops a lot from $14.3$ dB to $8.3$, indicating the effectiveness of the augmentation. We also present a NeuroHeed model (\#11) that is trained with SDR loss instead of SI-SDR. The performance also degrades a lot compared to system \#9.

\subsection{Online evaluation}
\label{sec:online}

\subsubsection{Comparison with offline evaluation}
In Table~\ref{tab:on_line}, systems \#8 and \#9 are the same trained models as in Table~\ref{tab:off_line}, but are used for online evaluation with a sliding window ($w_b+w_c$) instead of utterance-level inference. Both do not include the speaker encoder. The results show that offline systems consistently outperform their online counterparts. Comparing the offline-online pairs, the average SI-SDRi of system \#8 drops from $10.5$ dB to $7.8$ dB, due to the shorter EEG signal used in the sliding window. The average SI-SDRi of system \#9 has a higher degradation from $14.3$ dB to $8.3$ dB, due to the shorter EEG signal used in the sliding window, and also because it is trained in a non-causal way but evaluated in a causal way. In the latter, only the last part of the speech (window $w_c$) is used without any access to future frames.

\begin{figure}
  \centering
  \includegraphics[width=0.99\linewidth]{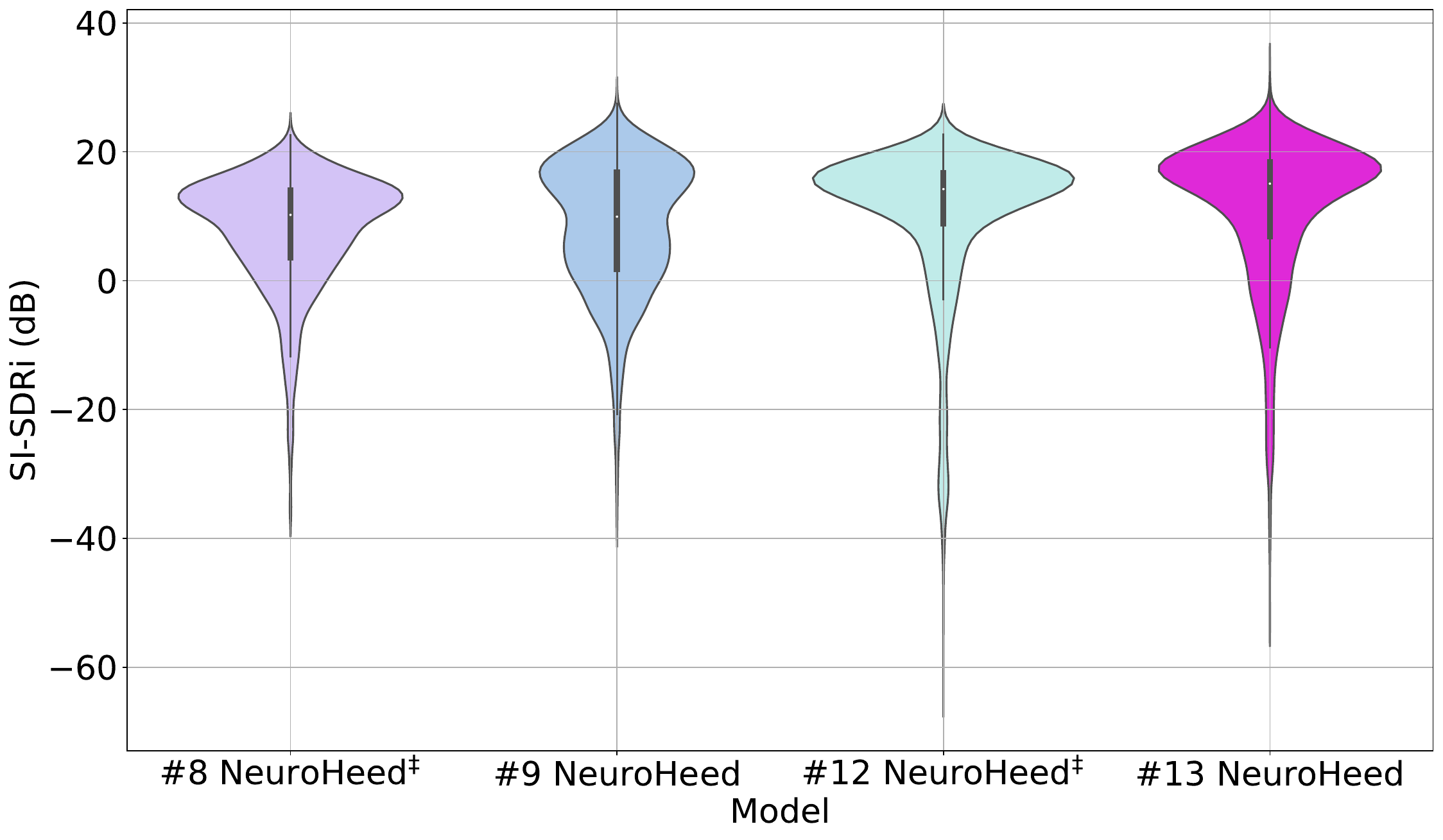}
  \caption{Violin plot of extracted speech signal's SI-SDRi from the test set for various models in online evaluation.}
  \label{fig:online_violin}
\end{figure}

\begin{figure}
  \centering
  \includegraphics[width=0.99\linewidth]{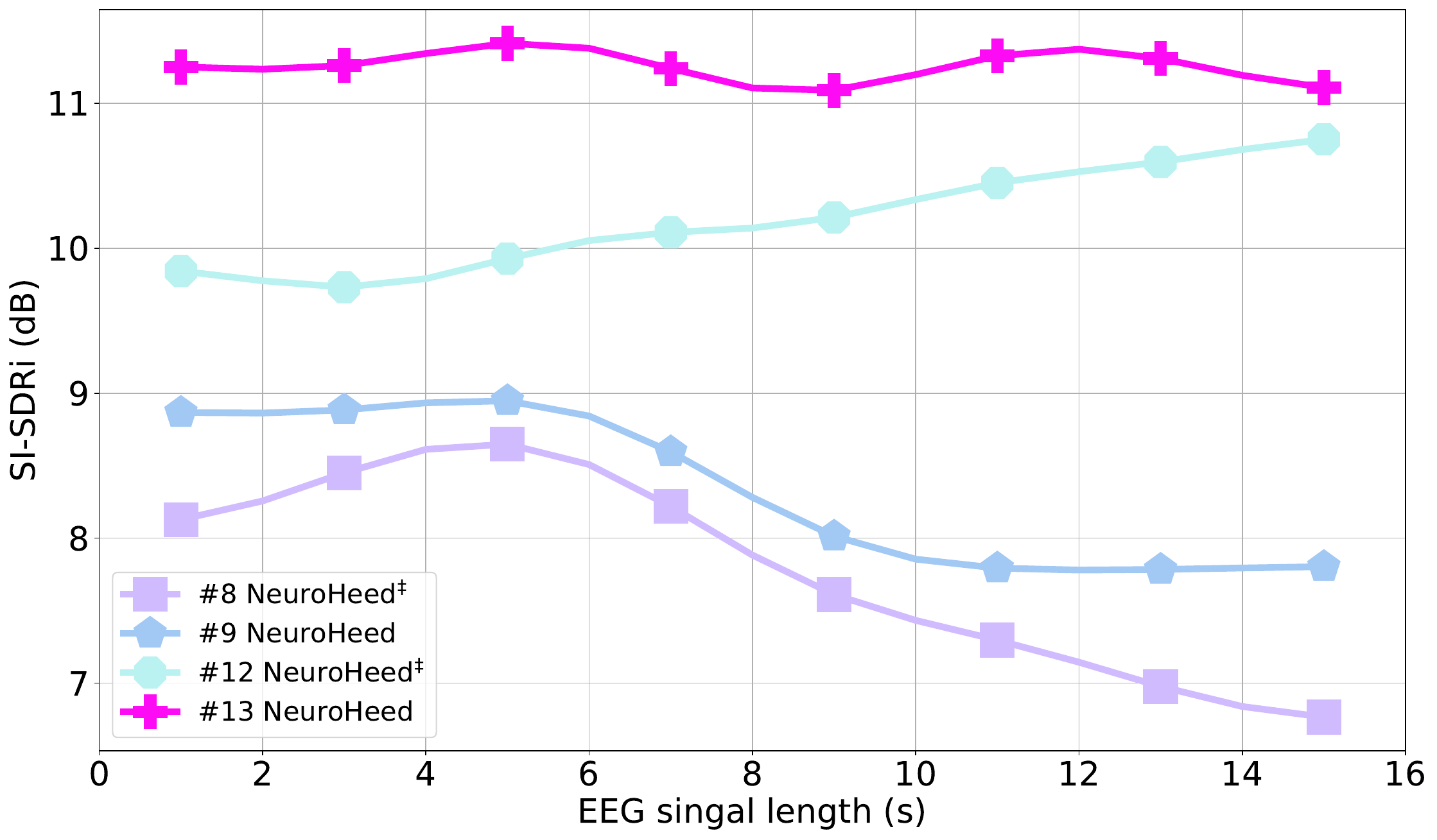}
  \caption{The average SI-SDRi for various utterance lengths in the online evaluation. Longer utterances mean longer EEG signals associated.}
  \label{fig:online_time}
\end{figure}

\begin{table*}[t]
    \centering
    % \sisetup{
    % detect-weight, % Make siunitx detect align bold cells correctly
    % mode=text, % Make siuntix print tables in text mode (causes width of bold characters to be the same as non-bold)
    % tight-spacing=true,
    % round-mode=places,
    % round-precision=2,
    % table-format=2.2
    % }
    \caption{Online evaluation for the EEG-based neuro-steered speaker extraction with different window lengths for $w_b$ in seconds. The window length for $w_c$ is set to 0.1 seconds. The average SI-SDRi is reported in dB.}
    % \addtolength{\tabcolsep}{-3.5pt}
    % \resizebox{\linewidth}{!}{
    \begin{tabular}{c|c|c|c c|cc|cc|cc|cc} 
       \toprule
        \multirow{2}{*}{Sys. \#}     &\multirow{2}{*}{Model}   &\multirow{2}{*}{Speaker encoder} &\multicolumn{2}{c|}{1 s}   &\multicolumn{2}{c|}{1.5 s}  &\multicolumn{2}{c|}{2.5 s}  &\multicolumn{2}{c|}{5 s}    &\multicolumn{2}{c}{10 s}\\
        &&  &SI-SDRi&RTF&SI-SDRi&RTF&SI-SDRi&RTF&SI-SDRi&RTF&SI-SDRi&RTF\\
        \midrule
        9   &\multirow{2}{*}{NeuroHeed} &\multirow{1}{*}{\xmark} 
            &4.0   &6.6    &6.2   &6.0    &8.3   &4.5  &10.0    &3.3   &10.4   &2.6    \\  
        13   &&\multirow{1}{*}{\cmark} 
            &7.7   &4.2    &9.7   &4.5    &11.2  &3.6  &12.0    &2.7   &12.0   &2.2    \\
        \bottomrule
    \end{tabular}
    % }
    % \addtolength{\tabcolsep}{3.5pt}
    % \vspace*{-3mm}
    \label{tab:online_wb}
\end{table*}

\begin{table*}[t]
    \centering
    % \sisetup{
    % detect-weight, % Make siunitx detect align bold cells correctly
    % mode=text, % Make siuntix print tables in text mode (causes width of bold characters to be the same as non-bold)
    % tight-spacing=true,
    % round-mode=places,
    % round-precision=2,
    % table-format=2.2
    % }
    \caption{Online evaluation for the EEG-based neuro-steered speaker extraction with different window lengths for $w_c$ in seconds. The window length for $w_b$ is set to 2.5 seconds. The average SI-SDRi is reported in dB.}
    % \addtolength{\tabcolsep}{-3.5pt}
    % \resizebox{\linewidth}{!}{
    \begin{tabular}{c|c|c|c c|cc|cc|cc|cc} 
       \toprule
        \multirow{2}{*}{Sys. \#}     &\multirow{2}{*}{Model}   &\multirow{2}{*}{Speaker encoder} &\multicolumn{2}{c|}{0.025 s}   &\multicolumn{2}{c|}{0.05 s}  &\multicolumn{2}{c|}{0.1 s}  &\multicolumn{2}{c|}{0.15 s}  &\multicolumn{2}{c}{0.2 s}\\
        &&  &SI-SDRi&RTF&SI-SDRi&RTF&SI-SDRi&RTF&SI-SDRi&RTF&SI-SDRi&RTF\\
        \midrule
        9   &\multirow{2}{*}{NeuroHeed} &\multirow{1}{*}{\xmark} 
            &7.5   &1.1    &7.9   &2.3    &8.3   &4.5   &8.3    &6.6    &8.7   &8.9\\  
        13   &&\multirow{1}{*}{\cmark} 
            &10.2  &0.9    &10.7  &1.8    &11.2  &3.6   &11.3   &5.4    &11.6  &7.2\\
        \bottomrule
    \end{tabular}
    % }
    % \addtolength{\tabcolsep}{3.5pt}
    % \vspace*{-3mm}
    \label{tab:online_wc}
\end{table*}

\subsubsection{Effect of online auditory attention attractor}
In online mode, we make use of the past extracted speech through an autoregressive speaker encoder. We study the effectiveness of this speaker encoder with systems \#12 and \#13 in Table~\ref{tab:on_line}. Comparing systems \#8 and \#12, the average SI-SDRi increases from $7.8$ dB to $10.1$ dB for the causal-TCN based NeuroHeed$^{\ddagger}$. In the violin plot of SI-SDRi shown in Fig.~\ref{fig:online_violin}, system \#8 has a flatter distribution compared to system \#12 and is centered around $14$ dB, while system \#12 has its center at around $16$ dB. System \#12 has some very negative samples at around $-30$ dB, which could be attributed to the speaker encoder providing misleading information due to poor past extraction performance. In Fig.~\ref{fig:online_time}, we plot the average SI-SDRi against the utterance length. As the utterances become longer, the average SI-SDRi of system \#8 increases a bit and then decreases, while the average SI-SDRi of system \#12 decreases a bit and then increases.

A similar trend is observed between systems \#9 and \#13, as shown in Table~\ref{tab:on_line}. The use of the speaker encoder increases the average SI-SDRi from $8.3$ dB to $11.2$ dB. Observed from Fig.~\ref{fig:online_violin}, system \#13 has more utterances centered at around $18$ dB while system \#9 has a flat distribution. In Fig.~\ref{fig:online_time}, as the utterances become longer, the average SI-SDRi increases a bit and then decreases for system \#9, while the average SI-SDRi of system \#13 remains stable.

We also show the violin plot of the SI-SDRi from individual subjects for the system \#13 NeuroHeed model in Fig.~\ref{fig:online_subject}. It can be seen that almost all subjects have a relatively wider distribution compared to the offline evaluation of system \#9 NeuroHeed model in Fig.~\ref{fig:offline_subject}. There is still a performance gap between the online and offline evaluations.

In addition, the DPRNN-based model has a higher RTF than the causal-TCN-based model due to the use of a large number of convolutional layers in the later model. In Table~\ref{tab:on_line}, system \#9 has a RTF of $4.5$ which is much better than system \#8 with $2.0$. System \#13 has a RTF of $3.6$ which is much better than system \#12 with $1.8$.

\subsubsection{Ablation studies}
As described in section~\ref{sec:norm}, we propose to evaluate the online models with a $1$ second initialization. In system \#13*, we evaluate the NeuroHeed model without this initialization. The average SI-SDRi drops from $11.2$ dB to $9.7$ dB compared to system \#13, showing that the $1$ second initialization is very important as the models perform worse when the sliding window is small.

In system \#13**, we do not apply the inference time normalization compared to system \#13. The average SI-SDRi drops from $11.2$ dB to $11.0$ dB, showing that normalization helps. We also present system \#14, which is trained using SDR as the loss function to retain the energy of the speech signals during extraction. It can be seen that the average SI-SDRi is only $8.4$ dB, far behind $11.0$ dB in system \#13**.

\subsubsection{Study on various sliding window lengths}
The online evaluation is done with a sliding window that is formed by the buffer signal window $w_b$ and currently acquired signal window $w_c$ as shown in Fig.~\ref{fig:window}. 

We first study the impact of various lengths of $w_b$ in Table~\ref{tab:online_wb}. It can be seen that system \#13 consistently outperforms system \#9 in terms of SI-SDRi for various lengths of $w_b$. For both systems, the RTF decreases as the length of $w_b$ increases because the models need to process a longer utterance for each sliding window. However, the performance in terms of SI-SDRi increases due to longer context information in the processing. System \#13 also outperforms system \#9 when they have a similar RTF, e.g., system \#13 has an average SI-SDRi of $9.7$ dB for a 1.5 s long window which outperforms system \#9 with $8.3$ dB for a 2.5 s long window. System \#13 has average SI-SDRi of $11.2$ dB for a 2.5 s long window which outperforms system \#9 with $10.0$ dB for a 5 s long window.

We also vary the length of the window $w_c$ and illustrate the results in Table~\ref{tab:online_wc}. The results indicate that system \#13 consistently outperforms system \#9 in terms of SI-SDRi for various lengths of $w_c$. For both systems, the RTF increases as the length of $w_c$ increases because fewer sliding windows are needed. The performance in terms of SI-SDRi increases slightly as the length of the perceived context increases marginally. However, the longer the length of $w_c$ is, the longer the delay in real-time inference, because the model needs to wait until the window $w_c$ of the speech signal is acquired to start the processing.

\section{Conclusion}
\label{sec:conclusion} 
In this paper, we proposed a time-domain end-to-end neuro-steered speaker extraction model, referred to as NeuroHeed, to perform selective listening based on the brain's elicited EEG signal. We modeled the EEG signals using a self-attention neural network and train it with the overall speech extraction quality as the objective function. The results suggest that the learned EEG feature representation is better than speech envelope reconstruction or TCN networks, as far as the detection of the brain's attention in speaker extraction is concerned. In the online NeuroHeed, we propose the autoregressive speaker encoder to self-enroll the attended speaker based on the past extracted speech that represents the attended speaker in the previous time frame. The self-enrollment offers less speaker confusion and higher signal quality in NeuroHeed.

\bibliographystyle{IEEEtran}
\bibliography{IEEEabrv,Bibliography}

\end{document}